\documentclass[5p]{elsarticle}
\usepackage{lineno,hyperref}
\modulolinenumbers[5]

\usepackage{booktabs} 
\usepackage{hyperref}
\usepackage{amsmath}
\usepackage{amsfonts}
\usepackage{amssymb}
\usepackage{chemarrow}
\usepackage{graphicx}
\usepackage{subfigure}
\usepackage{graphics}
\usepackage{bm}
\usepackage{rotating}
\usepackage{epsfig}
\usepackage[inline]{enumitem}
\usepackage{mathenv}
\usepackage{breqn}
\usepackage{url}
\usepackage{bm}
\usepackage{array}
\usepackage{algpseudocode}
\usepackage{lipsum}
\usepackage{xspace}
\usepackage{setspace}
\usepackage{tikz}
\usepackage{balance}

\newcommand*\cib[1]{\tikz[baseline=(char.base)]{
                            \node[shape=circle,fill=blue!50!black,text=white,draw,inner sep=0pt] (char) {#1};}}

\makeatletter
\usetikzlibrary{arrows}

{\bfseries}{\itshape}
{\bfseries}{\itshape}
{\bfseries}{\itshape}
{\bfseries}{\itshape}
{\bfseries}{\itshape}
{\itshape}

\newcommand*\bigcdot{\mathpalette\bigcdot@{.5}}
\newcommand*\bigcdot@[2]{\mathbin{\vcenter{\hbox{\scalebox{#2}{$\m@th#1\bullet$}}}}}

\newcommand{\note}[1]{}                                                 

\newcommand{\BfPara}[1]{{\noindent\bf#1.}\xspace\xspace}

\newcommand{\etal}{{\em et al.}\xspace}

\newcommand{\contraa}{{Contra-A}\xspace}
\newcommand{\contraf}{{Contra-F}\xspace}
\newcommand{\contras}{{Contra-S}\xspace}
\newcommand{\ours}{{Contra-*}\xspace}

\def\equationautorefname~#1\null{(#1)\null}

\begin{document}
\begin{frontmatter}
\title{Contra-*: Mechanisms for Countering Spam Attacks on Blockchain's Memory Pools}


\author[]{Muhammad Saad}
\ead{saad.ucf@knights.ucf.edu}
\author[]{Joongheon Kim}
\ead{joongheon@korea.ac.kr}
\author[]{DaeHun Nyang}
\ead{nyang@ewha.ac.kr}

\author[]{David Mohaisen\corref{mycorrespondingauthor}}
\cortext[mycorrespondingauthor]{Corresponding author}
\ead{mohaisen@cs.ucf.edu}
\fntext[myfootnote]{This paper is an extension of an earlier work that appeared in IEEE International Conference on Blockchains and Cryptocurrencies. (ICBC 2019)~\cite{SaadNKKNM19}}


\begin{abstract}
Blockchain-based cryptocurrencies, such as Bitcoin, have seen on the rise in their popularity and value, making them a target to several forms of Denial-of-Service (DoS) attacks, and calling for a better understanding of their attack surface from both security and distributed systems standpoints. In this paper, and in the pursuit of understanding the attack surface of blockchains, we explore a new form of attack that can be carried out on the memory pools (mempools), and mainly targets blockchain-based cryptocurrencies. We study this attack on Bitcoin's mempool and explore the attack's effects on transactions fee paid by benign users. To counter this attack, this paper further proposes \ours{}, a set of countermeasures utilizing fee, age, and size (thus, \contraf, \contraa, and \contras) as prioritization mechanisms. \ours optimize the mempool size and help in countering the effects of DoS attacks due to spam transactions. We evaluate \ours by simulations and analyze their effectiveness under various attack conditions.
\end{abstract}

\end{frontmatter}

\section{Introduction}\label{sec:introduction}

Blockchain technology promises to redefine trust in distributed systems by serving as a tamper-proof and transparent public ledger that is easily verifiable and difficult to corrupt~\cite{QuGLXYLZ20}. Blockchains use an append-only model backed by Proof-of-Work (PoW), augmenting trust in decentralized  Peer-to-Peer (P2P) settings. Due to such features, blockchains are used in cryptocurrencies, smart contracts, and Internet of Things (IoT)~\cite{AhmadSM19,NazaninAJ20}. The most widespread use of blockchains can be found in crytocurrencies, e.g., Bitcoin, Ethereum, Zcash, Litecoin, Altoins, etc.~\cite{GhoshGDK20}. 

Although blockchains are publicly verifiable and tamper-proof, they are vulnerable to attacks \cite{kaushal2016bitcoin,SaadSNKSNM20}. Moreover, since blockchains are most widely used in digital currencies, attackers have a high incentive to exploit them. In the last few years, attacks on cryptocurrencies have increased including including multiple 51\%\ (majority) attacks, selfish mining, double-spending, block withholding, block forks and distributed denial-of-service (DDoS) attacks \cite{sapirshtein2016optimal,rosenfeld2014analysis}.

In Bitcoin, DDoS attacks are launched against miners, users, and currency exchanges \cite{DDOS1}. In P2P systems, DDoS attacks may take various forms. For example, users or miners can be re-routed towards a counterfeit network, denying them access to the real network~\cite{SaadAAAYM19,SaadCLTM19}. Apostolaki~\etal \cite{apostolaki2017hijacking} estimate that an attacker can isolate more than 50\% of the network's hashing power by hijacking a few ($<$100) BGP prefixes. Moreover, attackers may exploit the block-size limit and network throughput to prevent benign users from getting their transactions verified. In Bitcoin, the average block size is limited to 1 MB and the average time of block mining is 10 minutes. On average, the transaction size varies from 200 Bytes to 1K Bytes.\footnote{We derive the average transaction size by dividing the Bitcoin block size by the total number of transactions in a block. These statistics are available at~\cite{BitcoinBlockchain20}} Under these constraints, Bitcoin can only verify three to seven transactions per second \cite{BanoAD17}, in contrast to mainstream payment processors such as Visa Credit which can verify up to 2000 transactions per second. Low transaction throughput creates a competitive environment where only selected transactions get accepted into a block. It also makes Bitcoin vulnerable to flood attacks \cite{baqer2016stressing}, where malicious users exploit the block size limit in Bitcoin ($\approx$1 MB)\footnote{The average Bitcoin block size is $\approx$1MB. Some miners use the SegWit protocol to increase the block size a little over 1MB. However, no Bitcoin block has exceeded a size of 1.4MB~\cite{BitcoinCommunity20,Canellis18}.} to overwhelm the blockchain with low-valued spam transactions. This further causes delay in verification of benign transactions. To prevent such attacks that exploit block size limit, miners in Bitcoin apply priority checks on the incoming transactions. The priority is given to the transactions that offer higher mining fee. Once high priority transactions are verified and mined into a block, the low priority transactions are queued into a memory pool.

In Bitcoin, memory pool (mempool) acts as a repository where all the transactions waiting to be confirmed are logged. Once a user generates a transaction, it is broadcast to the entire network. The transaction is stored into the mempool where it waits for confirmation. If the rate of incoming transactions at mempool is less than the throughput of the network (3-7 transactions/sec), there is no queue of unconfirmed transactions. Once the rate increases beyond the throughput, a transaction backlog starts at the mempool. Transactions that remain unconfirmed for long eventually get rejected. On November 11, 2017, the mempool size exceeded 115k unconfirmed transactions, resulting in USD 700 million worth of stall transaction  \cite{cryptocoinsnews_2017}. As the mempool size grows, users pay more mining fee per transaction to prioritize their transactions.

The Bitcoin network throughput is limited by the block size and the block mining time. Since the average block time is $\approx$10 minutes and the average block size is $\approx$1MB, Bitcoin can only process 3--7 transactions per second~\cite{BanoAD17}. Due to the limited throughput, users compete to get their transactions mined into the blockchain. Typically, users that pay a higher transaction fee win that competition. We note that the transaction fee is determined by the mempool size and if the mempool size is large, more transactions compete for the block. In this paper, we show that this competition can be exploited to launch a denial-of-service attack where an attacker can flood the Bitcoin mempool with unconfirmed transactions and inflate the mempool size. This attack makes the benign users believe that there is high competition in the mempool, and to win that competition, those users pay a higher mining fee~\cite{mempool_balooning}. We further show that in the current Bitcoin network, this attack can be easily launched since the default Bitcoin mempool does not apply any policy to filter the spam transactions. Therefore, our motivation is to identify the attack, show the attack feasibility, and propose the attack countermeasures

\BfPara{Contributions}  We make the following contributions. \cib{1} We identify the effect of mempool flooding on benign users in Bitcoin and the way that effect turns into a DoS attack.
\cib{2}  We present a threat model and associated attack procedure whereby an attacker can exploit the current Bitcoin protocol to achieve his goals.
\cib{3}  We propose \ours as countermeasures. \ours comes in fee-based (\contraf), age-bases (\contraa), and size-based (\contras), for transaction filtering. The three countermeasures optimize the mempool size, neutralize the attacker's capabilities, prevent mempool flooding, and put benign users at an advantage.
\cib{4} We examine the performance of our proposed countermeasures through discrete-event simulations and evaluate their performance under varying attack conditions. To the best of our knowledge, this is the first study  to address the problem of mempool attacks in cryptocurrencies with new mitigations. 

\BfPara{Organization} \autoref{sec:related} outlines the related work, and~\autoref{sec:preliminaries} outlines the preliminaries. In~\autoref{sec:threatmodel} and~\autoref{sec:attackprocedure} we describe the threat model and attack procedure that lead to mempool flooding and its associated effect. A modeling framework for analyzing \ours as well as evaluation metrics are introduced in~\autoref{sec:mod}. We propose countermeasures in~\autoref{sec:countering} with their associated analysis. Conclusion and future work are presented in~\autoref{sec:conclusion}.

\section{Related Work}\label{sec:related}
As described earlier, well-known attacks on blockchains include selfish mining, the 51\% attack, block withholding, double-spending, blockchain forks and DoS attacks. In this section, we review  notable work covering those attacks, and security aspects of blockchains.

Selfish mining is a form of attack where miners choose not to publish their block after computation, hoping to mine subsequent blocks and get more reward. The problem of selfish mining has been addressed by Eyal and Sirer~\cite{eyal2014majority} and Heilman~\cite{heilman2014one}. Eyal and Sirer~\cite{eyal2014majority}  proposed defense strategies to deter selfish mining attacks on blockchains. Block Withholding Attack (BWH), introduced in \cite{rosenfeld2011analysis}, is an attack in which miners in a pool choose to submit partial PoW, instead of the full proof. As a result, they get rewarded for participating in the pool although the pool suffers a loss due to partial solutions. Kwon \etal~\cite{kwon2017selfish} studied a new form of attack on blockchains called Fork After Withholding (FAW) attack which guarantees greater rewards than the block withholding attacks. 

The 51\% attack can be launched if a mining pool in the network gains more than 50\% of the network's hashing power. With more than half the hashing power of network, the attacker can prevent transactions from verification and other miners from computing a block.  To address the attack, the Two Phase PoW (2P-PoW) was proposed by  Eyal and Sirer~\cite{two_phased} and was analyzed by Bastiaan \cite{bastiaan2015preventing}. Double-spending or equivocation happens when a user generates two transactions from the same inputs and sends them to two recipients \cite{doublespending}. Double-spending can be countered by using one-time signatures in blockchains.

DoS attacks have been quite prevalent~\cite{vasek2014empirical}, and are repeatedly launched against the mining pools, benign users, and currency exchanges. Johnson \etal~\cite{Johnson2014} performed a game-theoretic analysis of DDoS attacks against Bitcoin mining pools. Vasek \etal~\cite{vasek2014empirical} illustrated DoS attacks empirically on the Bitcoin system. Cryptocurrency exchanges are frequently targeted to prevent coin tradings \cite{bitcoi16bg_2016}, and no mitigation to those attacks is proposed.

Another form of DDoS attack on blockchain includes spamming the network with low valued dust transactions. This attack is also called the {\em penny-flooding} attack. Baqer \etal~\cite{baqer2016stressing} performed Bitcoin stress testing to analyze the limitations of the Bitcoin network and how attackers exploit them. Similar to their work, in this paper we analyze the effect of penny-flooding attacks on users when a spam attack is carried out on the mempool of Bitcoin, and complement this analysis with countermeasures through memory pool optimization. To the best of our knowledge, this is the first study conducted to analyze the effect of spam attacks on mempool and explore their countermeasures.

\section{Preliminaries}\label{sec:preliminaries}

We now review the preliminaries of this work including details about the blockchain system and data collection. 

\begin{figure}[t]
\begin{center}
\includegraphics[ width=0.5\textwidth]{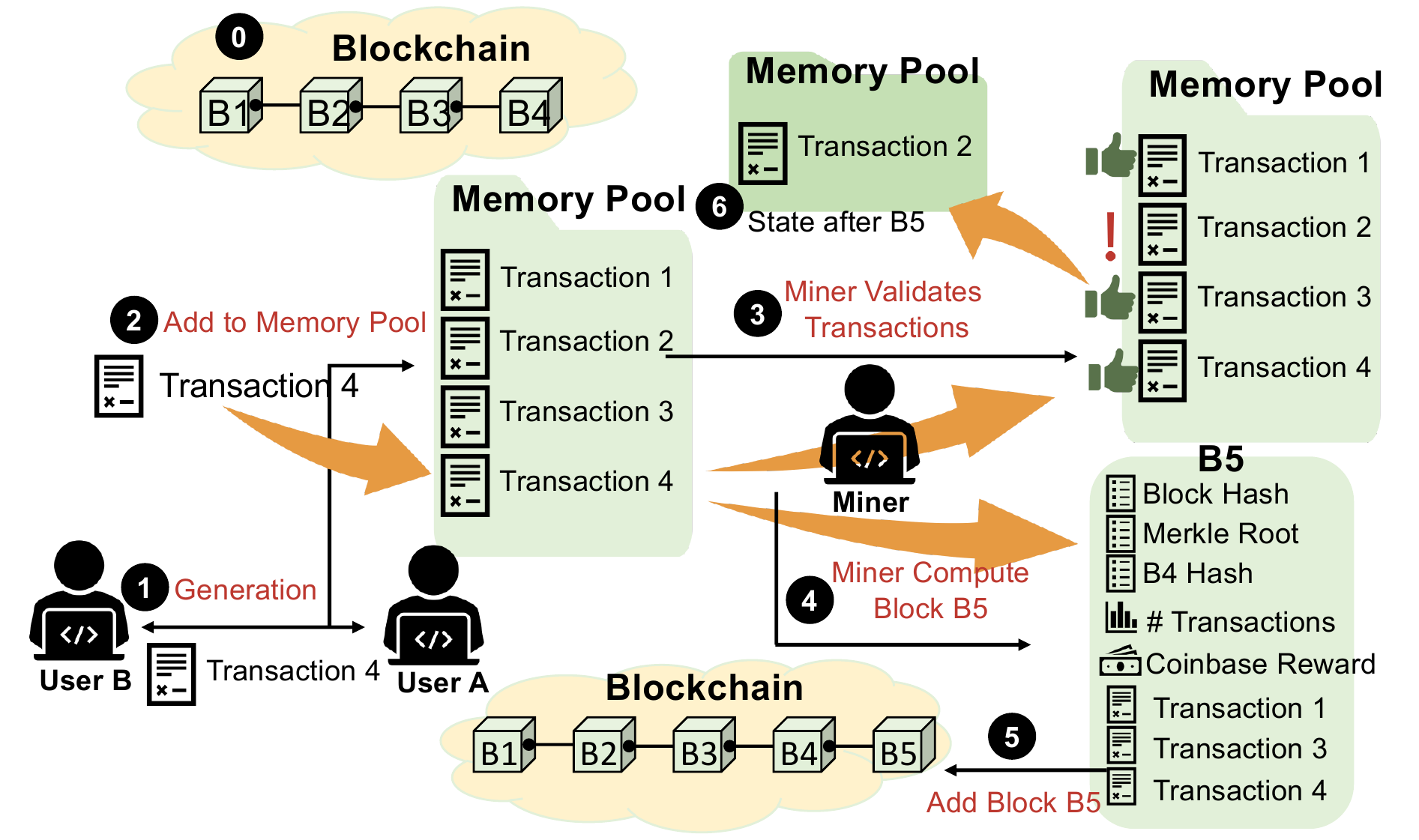}
\caption{Transaction lifecycle in a blockchain-based cryptocurrency. User A generates a transaction for user B. The transaction is stored in the memory pool along with other unconfirmed transactions. Miner validates transactions from memory pool, and computes a block. A valid block is added to the blockchain. }
\vspace{-3mm}
\label{fig:overview}
\vspace{-6mm}
\end{center}
\end{figure}

\subsection{An Overview of Bitcoin}
\BfPara{Transaction Lifecycle} Bitcoin users generate transactions and propagate them in the network. Transactions are temporarily stored in the mempool of Bitcoin nodes. Miners select transactions from the mempool and confirm them in a block. When a block is mined, it is released to the network and corresponding transactions in the mempool are removed.In~\autoref{fig:overview}, we provide an illustration of the transaction lifecycle in the Bitcoin network. 

\BfPara{Memory Pool} \label{parg:mempool} 
In cryptocurrencies, a memory pool (mempool) is the repository of unconfirmed transactions~\cite{bc-community}. As shown in~\autoref{fig:overview}, miners select transactions from the mempool and put them in a block. Typically, the mempool size is greater than the block size~\cite{BitcoinBlockchain20}, and the block size is limited to 1MB. If the mempool size exceeds the block size, it shows that the transaction arrival rate exceeds the confirmation rate. If the mempool size continues to grow, the transaction confirmation is delayed.  

\BfPara{UTXO} \label{parg:lifecycle} In Bitcoin, a user generates a transaction by using spendable balance in his wallet. The spendable balance includes transactions that are called ``Unspent Transaction Outputs'' (UTXO's)~\cite{SaadM18}. UTXO's are transactions received from other users and mined in the blockchain. For more details on the UTXO model, we refer to~\cite{bitcoindevelopers20}.

\BfPara{Relay Fee and Mining Fee} \label{parg:relayfee} In Bitcoin, the relay fee is the minimum fee a transaction must pay to be relayed by the network nodes. If a transaction does not pay the relay fee, it does not reach a miner's mempool. The mining fee is the fee paid to a miner as an incentive to confirm the transaction in a block \cite{bc-community}. Miners tend to prioritize those transactions that pay a higher mining fee. 

\BfPara{Confirmation} \label{parg:confirmation} Transaction confirmation means that (1) the transaction is successfully mined in a block, (2) the block is accepted by the network, and (3) the transaction is now a spendable UTXO in the receiver's wallet~\cite{bc-community}. A transaction confirmation score is the number of blocks mined after the block that includes the transaction. For example, if a transaction is mined at block height 10 and 5 blocks have been mined after that, the transaction confirmation score is 5. Note that the confirmation score is also called the age of the transaction. A 0 confirmation score means that either the transaction is not mined in a block or no block has been mined after the block that contains the transaction. Transactions that await mining are also called the ``unconfirmed transactions.''

\BfPara{Dust Transactions} \label{parg:dust} 
In cryptocurrencies, low paying transactions are called ``dust transactions'' \cite{kroll2013economics}, and they transfer a small value from the sender to the receiver. However, their transaction size is comparable to the size of a high-value transaction. Typically, attacks that target the block size limits are launched using dust transactions \cite{baqer2016stressing}.

\subsection{DDoS Attack on Mempools} \label{sec:ddosmem}
There are two types of DDoS attacks on blockchain-based cryptocurrencies. In the classical attack, the attackers exploit the block size limit by generating dust transactions. This attack was discovered by Baqer~\etal~\cite{baqer2016stressing}, and has been addressed by the Bitcoin community. The Bitcoin network applies a relay fee and a mining fee to filter dust transactions from the block. The second DDoS attack targets mempools by flooding them with unconfirmed transactions. Although these transactions may eventually be rejected by miners using countermeasures in \cite{baqer2016stressing}, their presence in the mempool creates another major problem. The mempool size determines the fee paid to the miners. If the mempool size is big, users compete to get their transactions mined by paying a higher fee. 

In \autoref{fig:Mempool}, we show a high correlation between the mempool size the transaction fee paid by users. Since the attacker's dust transactions get rejected, the attack does not affect his balance. In contrast, it invariably forces honest users to pay a higher mining fee. To the best of our knowledge, this work is the first attempt to notice this attack and propose countermeasures.  

\subsection{Data Collection}
\begin{figure}
	\centering
	\includegraphics[width=1\linewidth]{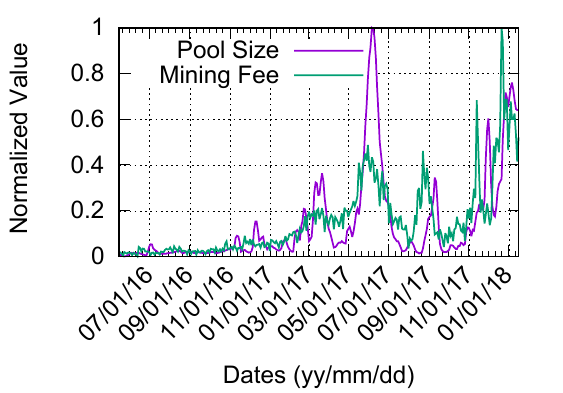}
	\caption{Temporal study of mempool size and mining fee paid by the users. Notice that as the mempool size grows, the mining fee increases accordingly. The spikes during May, September, and November indicate spam attacks.    }
	\label{fig:Mempool}
\end{figure}

To observe the relationship between the mempool size and the mining fee, we used the public dataset provided by the company called ``Blockchain''~\cite{BlockchainI18-tf}. For this study, we gathered the dataset of mempool size and fee from January 2016 to May 2018. In~\autoref{fig:Mempool}, we plot the results obtained from the dataset and we use the min-max normalization to scale our dataset in the range [0--1]. Our data shows that Bitcoin mempool has been attacked three times in 2017, resulting in an increased mining fee.

\section{Threat Model} \label{sec:threatmodel}
In this work, we assume that the attacker is a full Bitcoin client with a complete blockchain and a memory pool at his machine. The attacker's wallet has spendable bitcoins denoted by ``UTXO's''; those bitcoins have been previously mined into the blockchain. In Bitcoin, transactions can be split into various small transactions \cite{nakamoto2008bitcoin}. We assume that the balance in the attacker's wallet is large enough that it can be split into fractions of dust transactions; each of those transactions is at least capable of paying the mining fee. We assume also that the attacker controls a group of sybil accounts, each with multiple public addresses. These public addresses could be used to exchange transactions during an attack. The attacker and the sybil accounts have an apriori knowledge of each others' public addresses. Furthermore, the attacker and sybils have client side software and scripts \cite{caetano2015learning,bitcoinjs}, which enable them to initiate a flood of ``raw transactions'' \cite{rawtransactions} in a short time span. We assume that Sybils (collectively) have a capacity of exchanging transactions at a much higher rate than the network's throughput \cite{croman2016scaling}.

Although, being a full client in the network, we assume that the attacker does not have the capability of mining transactions. This means that the attacker does not possess enough computational power to mine a block, discard a transaction, reverse a transaction or delay other transactions from being mined. Moreover, the attacker does not have control over other benign users in the network and, as such, cannot prevent them from broadcasting their transactions and accessing mempool nor other resources in the network. The attacker is also constrained by a ``budget'': since every transaction requires a minimum fee to be relayed to the network, the fee limits the number of transactions that an attacker can generate.

\BfPara{Attacker's Goal} The end goal of the attacker is to flood  mempools in the network with {\em dust transactions}. The attacker will broadcast dust transactions at a higher rate than the throughput of the network. At mempools, the arrival rate corresponds to the flow of incoming transactions and the departure rate corresponds to the rate of transaction mining. The departure rate is fixed, because the average block computation time and the size of the block are currently fixed. When the arrival rate increases due to a flood of dust transactions, it results in transactions backlog. As the queue size grows, the mempool size increases accordingly. Overwhelming the mempool size alarms the benign users, who naturally start paying higher mining fee to prioritize their transactions. 

Upon flooding, a secondary objective of the attacker is to reduce the cost of attack by causing the mempool to reject his transactions. For the attacker, mining transactions will result into losing the associated mining fees to the miners. However, if the transactions get rejected, the attacker will have another chance to launch attack. 

In the prior work, the target of attack was either a mining pool \cite{Johnson2014}, a Bitcoin exchange \cite{wueest2014continued} or the blockchain itself \cite{baqer2016stressing}, leading to various avenues of DoS attacks. In our work, on the other hand, the target of the attack is the mempool of the system outlined in~\autoref{fig:overview}, while the potential victims in every attack are the benign users in the blockchain network who are denied service. Another distinguishing feature of this threat model is that the attacker does not want his transactions to be mined. In the analysis performed by Baqer \etal \cite{baqer2016stressing}, the intent of the spam attack was to flood blocks in the blockchain by exploiting the block size limit. Such an attack requires transaction mining in the blockchain. In Bitcoin, the gap between the block size and the mempool size presents an opportunity that can be exploited for an attack.\footnote{By default, the mempool size at each Bitcoin node is 300MB. Since the average block size is limited to $\approx$1MB, the gap between the block size and mempool size can be exploited to flood the network with dust transactions. The mempool size can be adjusted in the {\em Bitcoin.conf} file. However, decreasing it to the average block size may result in losing valuable transactions for mining~\cite{BCommunity20,bitcoin_2020}} Note that mempool flooding in the network does not require require transaction mining. Moreover, attacks exploiting block size can be effectively countered by miners while the attack on mempools cannot be countered in the same way. As transaction mining involves mining fee while the attack objective is only mempool flooding. Therefore, transaction acceptance in a block remains undesirable for the attacker in this attack. According to the taxonomy of DoS attack \cite{mirkovic2004taxonomy}, the mempool flooding attack can be characterized as a ``semantic attack of variable rate.''

\begin{figure*}
\centering
\includegraphics[width=1\linewidth]{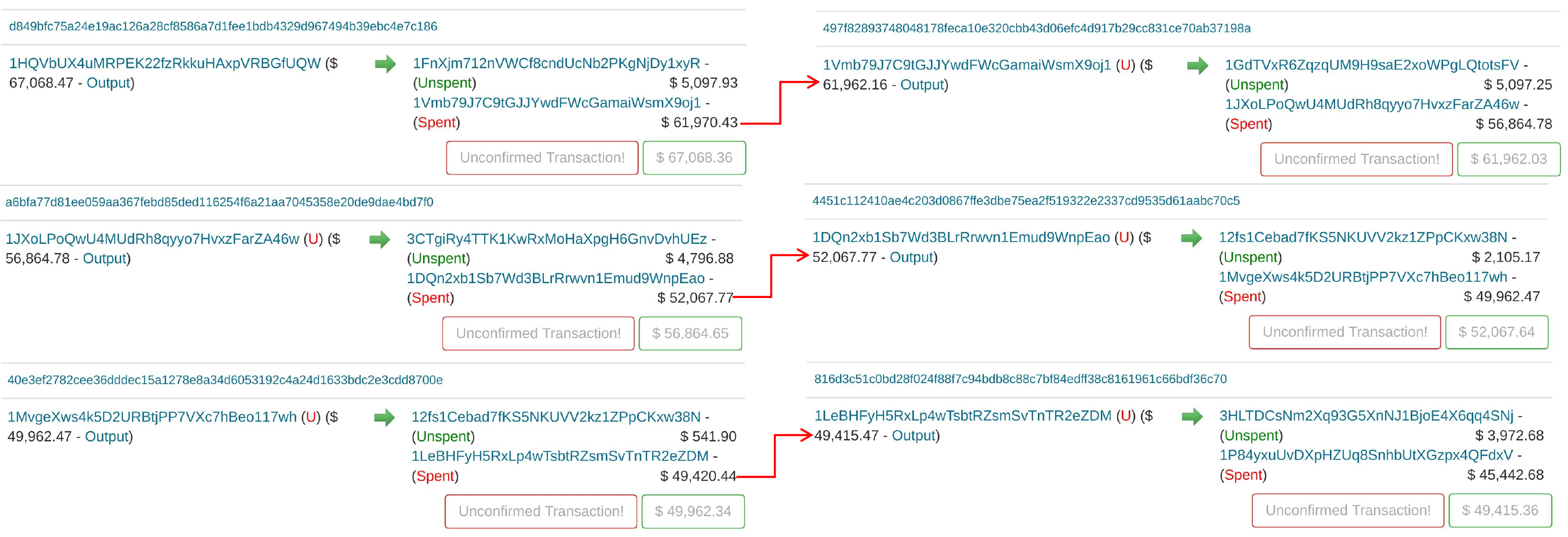}

\caption{Sequence of transactions obtained from blockchain.info on May 3rd, 2018. Notice that one of the addresses as an output is used as the input address to the next transaction. The difference in the output balance and input balance is the fee paid by the transaction. The figure shows that unconfirmed transactions can be spent prior to confirmation and while they may not be selected by the miners, they still overwhelm the mempool. This depicts the attack model where Sybils exchange unconfirmed transactions at a high rate. } 

\label{fig:usp}

\end{figure*}

\section{Attack Procedure} \label{sec:attackprocedure}

As mentioned earlier, when the rate of incoming transactions exceeds the network's throughput, a backlog of unconfirmed transactions builds up at the mempool. As backlog grows, competition for transaction mining also increases. Users try to prioritize their transactions by offering more fee to the miners. As a result, the fee per transaction paid to the miners increases. To facilitate usage, there are online services such as ``Bitcoin Fee Estimation'' \cite{estimatefee}, which estimate mempool size and the average fee paid per transaction. Therefore, a fee is recommended for users who want their transactions confirmed within desired time. 

Since the mempool size affects the way users pay the mining fee, this creates an attack possibility for an attacker to exploit the mempool size and create panic among benign users. When a benign user sees the mempool size growing, the user, as a rational agent, will try to prioritize his transactions by adding more mining fees to them. Dust transactions of an attacker will eventually be rejected by the miners, to protect blockchain from spam. Although it protects the system from spam, that policy in itself also works in favor of the attacker, since the attacker loses no bitcoins as a result of enforcing this policy. On the other hand, benign users end up paying more than the required fee to get their transactions confirmed. Upon rejection, the attacker can re-launch the attack multiple times.

As shown in~\autoref{fig:Mempool}, there is a high correlation between the mempool size and the transaction fee paid to the miners. In May, August and November 2017, it was reported~\cite{spamattack_bitcoin7} that Bitcoin was under spam attack of unconfirmed transactions which led to higher mining fee. From~\autoref{fig:Mempool}, it can be observed that during those months the size of the mempool was much larger than the average size. As a result, the mining fee pattern also followed similar growing trend as the mempool size. In December 2017, the problem of mempool flooding with was highlighted by crypto analysts~\cite{Young_18} suggesting that it was an attempt to increase the mining fee and drive the users away from Bitcoin. To further establish the relationship between the mining fee and the mempool size, we computed Pearson correlation on our dataset. The Pearson correlation coefficient is defined as
$\rho(X,Y) = \frac{{\bf
Cov}(X,Y)}{\sqrt{{\bf Var}(X){\bf Var}(Y)}}$, and from our dataset, we observed a high correlation of 0.69 between the mempool size and the mining fee.  

As a result, we conclude that overwhelming the mempool size can also lead to other problems in the blockchain. In November 2017, when the mempool was flooded, USD 700 million worth of bitcoins remained stuck in the pool for two days~\cite{cryptocoinsnews}. Delay in verification can create multiple problems, including possibilities of double-spending.

As described in the threat model, the objective of the attacker is to maximize the size of the mempool and minimize the cost of the attack. The cost of the attack is the fee paid to the miner if a transaction gets mined in a block. The fee consists of the relay fee\footnote{The relay fee is a cap implemented by nodes to filter spam transactions. The relay fee threshold can be adjusted in the {\em Bitcoin.conf} file~\cite{relay_fee}. Typically, if a node receives a transaction with 0 fee, the node discards the transaction and does not forward it to other nodes. Therefore, it protects the network from spam transactions. Realizing this utility, we consider the relay fee for our model. } and the mining fee. Higher fee increases the priority of transaction and chances of a transaction mining. To avoid that, the attacker will design his transactions in a way that they are less likely to be prioritized by miners. At the same time, the attacker wants his transactions to stay in the mempools for as long as possible. As such, this attack can be carried out in two phases: the distribution phase and the attack phase.  

\subsection{The Distribution Phase} \label{sec:firstphase} In the distribution phase, the attacker estimates the minimum relay fee of the network, divides his spendable bitcoins (``UTXO's'') into various transactions and sends them to the sybil accounts. This can be done in two ways. 1) The attacker may generate a dust transaction from previous UTXO, send it to a sybil wallet and get the change back as a new transaction. The attacker uses the change as new input balance and repeats the procedure multiple times for all sybil addresses. 2) An alternative way is to use the spendable balance and generate a series of outputs to all the addresses of sybil nodes. Unlike the previous method, this will result in only one transaction to all the sybil outputs. Transactions of this nature are known as ``sendmany'' transactions~\cite{sendmany}, because the user is sending bitcoins to various addresses within one transaction. Since the aim of the attacker is to generate as many transactions as possible, he will not opt for ``send many'' option. All transactions to the sybil wallets will be generated independently. The transactions made in the distribution phase will have input ``UTXO's'', which will be previously mined in the blockchain. Hence, these transactions will have greater-than-zero age (confirmation score), and will be capable of paying the minimum mining fee.

\subsection{The Attack Phase}\label{sec:secondphase} 
Once the distribution phase is completed, all sybil accounts will have sizable balance in their wallets. In the attack phase, all sybils will carry out ``raw transactions'' \cite{rawtransactions} from the balance received in the distribution phase. Sybils will generate dust transactions and exchange them with each other. To maximize the severity of attack, they will prefer to have one recipient per transaction. The rate of transactions will be much higher than the throughput of the network. As a result, the arrival rate of the transactions at the mempools will be higher than the departure rate of mined transactions. This will increase the transaction backlog and the size of the mempools over the duration of the attack. The attack will be carried out until all the spam transactions get into the mempools. The transactions made in the attack phase will have the transactions of distribution phase as input ``UTXO's''. These inputs will still be awaiting confirmation in the blockchain. As such, their confirmation factor or age score will be zero.

\subsection{Attack Illustration} \label{sec:illus}
In \autoref{fig:usp}, we provide an illustration of such an attack whereby unconfirmed transactions were exchanged. The letter ``U'' (in red) shows the unconfirmed transaction. As shown in~\autoref{fig:usp},  various transactions were generated to flood the mempool within a short time duration. The user started with one confirmed transaction in the blockchain. The transaction was later split into two transactions (the distribution phase). Out of the two resulting transactions, one was immediately spent. Moreover, the spent transaction was unconfirmed, thus the letter ``U'' encoded (in red) was used to denote its status. Among all the transactions exchanged, each transaction paid a fee of 8.27 USD, 0 USD, and 4.97 USD, respectively. Compared to the value exchanged in each transaction, the percentage of fee was 0.01\%, 0\%, and 0.01\%. All these transactions had zero confirmation score. The attack phase in depicts a similar case where sybils exchange such transaction at a high rate.

\subsection{Attack Cost} \label{sec:cost}

As mentioned earlier, one of the objectives of attacker is to minimize the attack cost. To be able to achieve that, the attacker requires transactions to be part of the mempool but not part of the blockchain.  This can be achieved by adding the minimum relay fee ($R_f$) to each transaction but not the minimum mining fee ($M_f$). The relay fee is necessary for a transaction to be broadcast to all peers in the network and be accepted by the mempool.  If the attacker adds the mining fee, his transactions will attain priority from a miner and might  get mined. To avoid that, the sybils only pay the relay fee. If a transaction has $i$ inputs, where each input contributes a size of $k$ Bytes, and $o$ outputs, where each output contributes a size of $l$ Bytes, then the total size of the transaction $S$ and its associated cost $C$ are determined by~(\ref{cost:size})-(\ref{cost:one}), respectively.

\begin{align} 
 \label{cost:size} S (\text{Bytes}) &=  (i \times k) + (o \times l) + i\\
  \label{cost:one}C (\text{BTC}) &= R_f\times\frac{S}{1024} = R_f \times\frac{[(i \times k) + (o\times l) + i ]}{1024}
\end{align}

Assuming that the attacker is limited by a budget $B$ (BTC) and minimum transferable value set by the network as $T_\min$, then, using~\autoref{cost:one}, the total number of transactions  $T_a$ that the attacker can generate can be computed in~\autoref{equation:total_transactions}. At the time of writing of this paper, the minimum transferable value in Bitcoin was 5460 Satoshis \cite{minimum_fee}.

\begin{equation} \label{equation:total_transactions}
    T_a  = \frac{B \times 1024}{R_f \times T_\min\times[(i \times k) + (o \times l) + i ] }
\end{equation}

Now we look at the system from standpoint of a benign user. A benign user who intends to get his transaction mined into the blockchain pays relay fee for transaction broadcast and mining fee as an incentive to the miner. For such a user, contributing a total $T$ transactions, the cost incurred per transactions and the total cost of all transactions $T_l[R_f + M_f]$ can be derived using~\autoref{equation:legit} and~\autoref{equation:costlegit} .
 \begin{align} \label{equation:legit}
    C (\text{BTC}) &= [R_f + M_f]\times\frac{[(i \times k) + (o \times l) + i ]}{1024} \\
\label{equation:costlegit}T_l (\text{BTC}) &= \textit{T} \times  [R_f + M_f] \times \frac{[(i \times k) + (o \times l) + i ] }{1024}
\end{align}

As mentioned in the threat model (\autoref{sec:threatmodel}), the aim of the attacker is to increase the cost per transaction paid by the benign user~\autoref{equation:legit}. A benign user will aim to have his transactions mined, therefore he will pay relay fee and a high mining fee. The attacker will only aim to get his transactions into the mempool and eventually not get mined, so he will only pay the relay fee. In these settings, the maximum loss an attacker can incur would happen if all his transactions get mined. The cost in such a case will be equal to the  product of the total number of transactions and the relay fee ($T_a\times R_f$). The attacker can re-launch the same attack with a new balance of $B-(T_a\times R_f$). If a portion of the attacker's total transactions $t_a$ gets mined, where $t_a \leq T_a$, then the attacker would be able to re-launch the attack with new balance of  $B-(t_a\times R_f)$. 

\subsection{Attack Feasibility}\label{sec:af}
Our threat model and attack procedure are feasible in the real-world Bitcoin network. As shown in~\autoref{sec:cost}, the minimum transferable value in Bitcoin is 5460 satoshis which is equal to 546$\times10^{-7}$ bitcoins~\cite{minimum_fee}. If the attacker has 1 bitcoin, he can generate up to 18315 transactions at any time. Since the Bitcoin network throughput is 3--7 transactions per second, therefore, to launch an effective attack, the attacker needs to generate transactions at a faster rate than the network throughput. Assuming the upper bound throughput limit (7 transactions per second), the attacker can sustain the attack for (18315/7) $\approx$2616 seconds. Since the average Bitcoin block time is 600 seconds, Therefore, with only 1 bitcoin, the attacker can target up to four consecutive blocks.

Moreover, since all transactions are dust transactions, they are less likely to be mined in a blockchain and will stay in the mempool for a long time. If those transactions are evicted from the mempool, the attacker can reissue them. As a result, the attack will persist even after four blocks and benign users will continue to pay a higher transaction fee. Also note that the attacker can control all Sybil accounts from a single machine since multiple wallet addresses can be generated from a single blockchain node~\cite{BitcoinBlockchain20}. Therefore, the attack does not require the adversary to control multiple machines.

From the above analysis, we make the following key conclusions. (1) The adversary does not require a significant balance to launch the mempool DDoS attack since only 1 bitcoin can suffice. (2) With only 1 bitcoin, the adversary can easily cross the maximum transaction throughput to inflate the mempool size. (2) Given that the attack is less costly and can be easily managed from a single blockchain node, a single attacker can launch the attack instead of multiple attackers jointly targeting the network. All these factors justify our threat model, making the attack more practical in the real-world Bitcoin network.

\section{Modelling The Mempool Attack} \label{sec:mod}
As mentioned in \autoref{sec:preliminaries}, mempool acts as a buffer for unconfirmed transactions, where the incoming transactions denote the arrival, and transaction mining represents the departure process. As long as the arrival process is within the bounds of system's throughput (3--7 transactions per second), the mempool queue remains stable and there is no transaction backlog. However, as shown in our threat model, the attacker overflows the buffer by accelerating the arrival process and increasing the queue size. Therefore, to construct effective countermeasures, it is useful to formulate this abstraction as a queuing theory problem with necessary mathematical primitives. To that end, we model the mempool attack as Lyapunov optimization problem that encapsulates the attack procedure and provides a roadmap towards the attack countermeasures.

\subsection{Lyapunov Optimization} \label{sec:lyp}
Lyapunov optimization is a popular scheme applied to the field of dynamic control systems for time-average optimization under stability constraints \cite{QiuHCZ18}. In queuing networks, Lyapunov drift is used to model the queues while optimizing time-average performance objectives such as the energy or the throughput. Subject to the queue stability, i.e., $\lim_{t\rightarrow\infty}\frac{1}{t}\sum_{\tau=1}^{t-1}Q[t]<\infty$ where $Q[t]$ is the queue backlog at $t$, the time-average Lyapunov optimization (i.e., drift-plus-penalty (DPP)~\cite{tii06neely,ton16kim}) is defined as:
\begin{align} 
\label{eq:lp1} Q_{o}(t+1) &\geq Q_{o}(t) + y_{o}(t)\\
\label{eq:lp2} y_{o}(t) &\leq Q_{o}(t+1) - Q_{o}(t)\\
\label{eq:lp3} \sum_{\tau = 1}^{t-1}\nolimits y_{o} [\tau] &\leq Q_{o}(t+1) - Q_{o}(0) = Q_{o}(t)  \\
\label{eq:lp}\textbf{Minimize}  &: \lim_{t\to\infty} \frac{1}{t} \sum_{\tau = 1}^{t-1}\nolimits y_{o} [\tau]
\end{align} 
where $y_{o}[\tau]$ is the objective function at $t$. Note that $Q[t]$ is a discrete time queue where $Q_{o}[t]$ is the queue backlog at any time and $y_{o}[\tau]$ is the difference between arrivals and departures on the given time slot. A stable queue ensures that the time average of the objective function is minimum. Based on DPP, this time-average optimization subject to queue stability can be formulated as the following decision-making framework:

\begin{figure}
\centering
\includegraphics[width=1\linewidth]{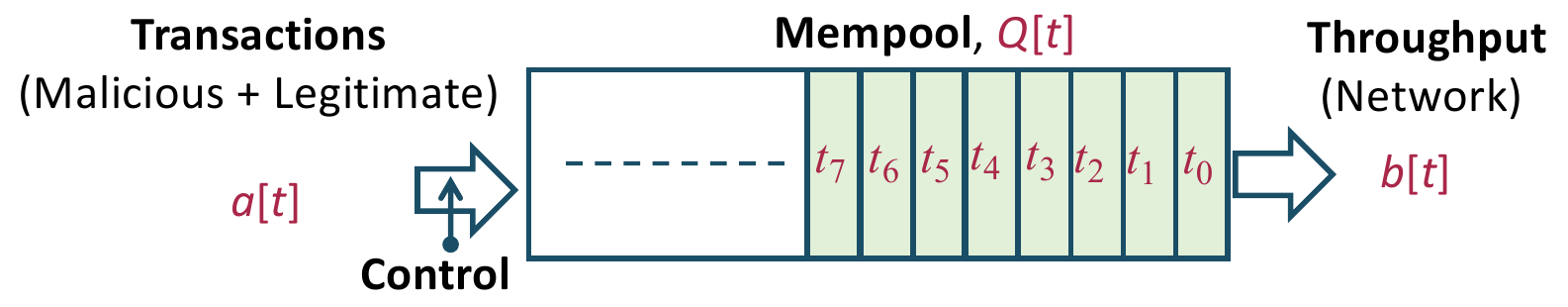}
\caption{Mempool as a queue with arrival and departure processes. The arrow with control is where we apply optimization techniques. } 
\label{fig:queue}
\end{figure}

\begin{equation}
\label{eq:lpm}
\alpha^{*}[t]\leftarrow \arg\min_{\alpha[t]\in\mathcal{A}}\left[V\cdot y_{o}[\alpha[t]]+Q[t]\left\{a(\alpha[t])-b(\alpha[t])\right\}\right], 
\end{equation}
where $\alpha[t]$ is possible decision at $t$, $\alpha^{*}[t]$ is time-average optimal decision at $t$, $\mathcal{A}$ is a set of possible decisions, $V$ is tradeoff coefficient between optimization criteria and stability, 
$y_{o}[\alpha[t]]$ is objective value with decision  $\alpha[t]$ at $t$,
$a(\alpha[t])$ is an arrival with decision $\alpha[t]$ at $t$, $b(\alpha[t])$ is a departure with decision $\alpha[t]$ at $t$, respectively. 

To apply this scheme to the mempool attack, we first model the mempool as a queue and show the arrival and departure processes. We present our design model in \autoref{fig:queue}, where the arrival process consisting of benign and malicious transactions is denoted by $a[t]$, the departure process is denoted by $b[t]$, and the mempool is denoted by $Q[t]$. Since the mining rate eventually determines the throughput of the network, therefore, we use throughput as the departure process. The control shown in the red allows us to apply countermeasures and modify the queue size. Our objective is to minimize the time-average size, subject to the queue stability. Queue stability can be achieved by removing malicious transactions or limiting the total number of transactions in the queue. The cost is the unwanted removal of benign transactions as an outcome of the applied countermeasure. Therefore, we have two cases that represent the state of the mempool. In the first case, when the mempool is idle, the applied control policy should only remove a small number of transactions to fulfill the objective function of time-average cost minimization. In the second case, when mempool is flooded, the applied control policy should remove a large number of transactions to guarantee mempool size stability. With these objectives, the Lyapunov-based DPP changes to \autoref{eq:lpm} is updated to \autoref{eq:lpf}, where $C (\alpha[t])$ is power consumption when our decision is $\alpha[t]$. If the mempool is empty ($Q[t] = 0$), we have to minimize our cost by permitting more arrivals. If the mempool is flooded ($Q[t] \approx \infty$), we have to minimize $(\alpha[t])$ by removing malicious transactions and stabilizing the mempool size. 
\begin{align} 
\label{eq:lpf}
\alpha^{*}[t]\leftarrow \arg\min_{\alpha[t]\in\mathcal{A}}\nolimits
\left\{V\cdot C[\alpha[t]]+Q[t]a(\alpha[t])\right\}. 
\end{align} 

The formal attack model provides us the following key insights that can be used to develop effective countermeasures for the DoS attack: 
\begin{enumerate*}
    \item The mempool acts as a buffer for incoming transactions. Therefore, the countermeasure policy should not only reduce the number of malicious transactions but also ensure the queue stability to ensure operational consistency. 
    \item Since the network throughput is constant, therefore, we cannot control the rate of outgoing transactions. As such, our countermeasures must be applied at the control knob of the queue. 
\end{enumerate*}

\subsection{Evaluation Parameters} \label{sec:evalpm}
We use the following parameters to examine effectiveness of \ours; \contraf, \contraa, and \contras. 

\BfPara{Precision} Precision measures the relevant information obtained from an experiment with respect to the total information. It can be computed as the ratio of true positive and the sum of true positive and false positive $\frac{TP}{TP+FP}$. 

\BfPara{Recall} Recall is the measure of relevant information obtained from an experiment with respect to the total relevant information. Mathematically, it is defined as the ratio of true positive and the sum of true positive and false negative $\frac{TP}{TP+FN}$. 

\BfPara{F1 Score} F1 score uses both precision and recall and provides their harmonic average. F1 score can be computed as $\frac{2\times precision \times recall}{precision + recall}$

\BfPara{Accuracy} In machine learning, accuracy measures the the classifier's strength in determining the experimental outcomes, computed as  $\frac{TP+TN}{TP+TN+FP+FN}$. 

\BfPara{Negative Rate} Negative rate or specificity is the measure of truly identified negatives in the complete set of negative values. Negative rate can be computed as $\frac{TN}{TN+FN}$.

We will use this evaluation criteria for all the experiments described in the rest of the paper.

\section{Countering The Mempool Attack} \label{sec:countering}

To counter DoS on Bitcoin's mempool, we propose three countermeasures that prevent the spam on the system. The design motivation is to increase the attack cost for the attacker while keeping the operational efficiency for benign users. We develop our designs for a miner's priority and apply the priority check on transactions at the mempool level. One of the effective countermeasures against spam attack in Bitcoin is to prevent the transmission of dust transactions in the network. We envision that if mempools can discard spam transactions and stop relaying them to other mempools, the pool size can be effectively controlled and spam can be countered.

\subsection{\contraf: Fee-based Mempool Design} \label{sec:first_policy}
Fee-based design aims to optimize the size of mempool by filtering spam transactions upon arrival. As the threat model states, an attacker only intends to relay spam transactions between the mempools and does not want them to be mined. To achieve this goal, the attacker only pays minimum relay fee in transactions so that mempools accept and relay them. To prevent the transactions from mining, the attacker does not pay the mining fee. We use this insight to construct a ``Fee-based Mempool Design.''

For this design in Algorithm 1, we assume that the mempool is initially empty when transactions begin to arrive. We also assume that each incoming transaction has its associated relay fee and mining fee. We also fix a threshold beyond which the mempool starts spam filtering. Initially, when the transactions arrive in the pool, and for each transaction, the mempool checks if the transaction pays a minimum relay fee. If the transaction pays the minimum relay fee, it is accepted and the mempool size is updated. As the transactions get added into the mempool, the size of mempool grows. When the size reaches a threshold, the mempool starts applying the fee-based policy. Now, if the incoming transaction pays both the minimum relay fee and the minimum mining fee, only then it is accepted in the mempool. The key idea behind this scheme is that only those transactions should be accepted, which eventually get mined into the blockchain. As a result, this technique puts a cap on the incoming transactions and filters spam transactions, thereby reducing mempool size. If the new size is less than the baseline size threshold then the mempool can proceed its operation from relay fee check. Otherwise, it will continue with the fee-based design.

\begin{figure}
\centering
\includegraphics[width=0.5\textwidth]{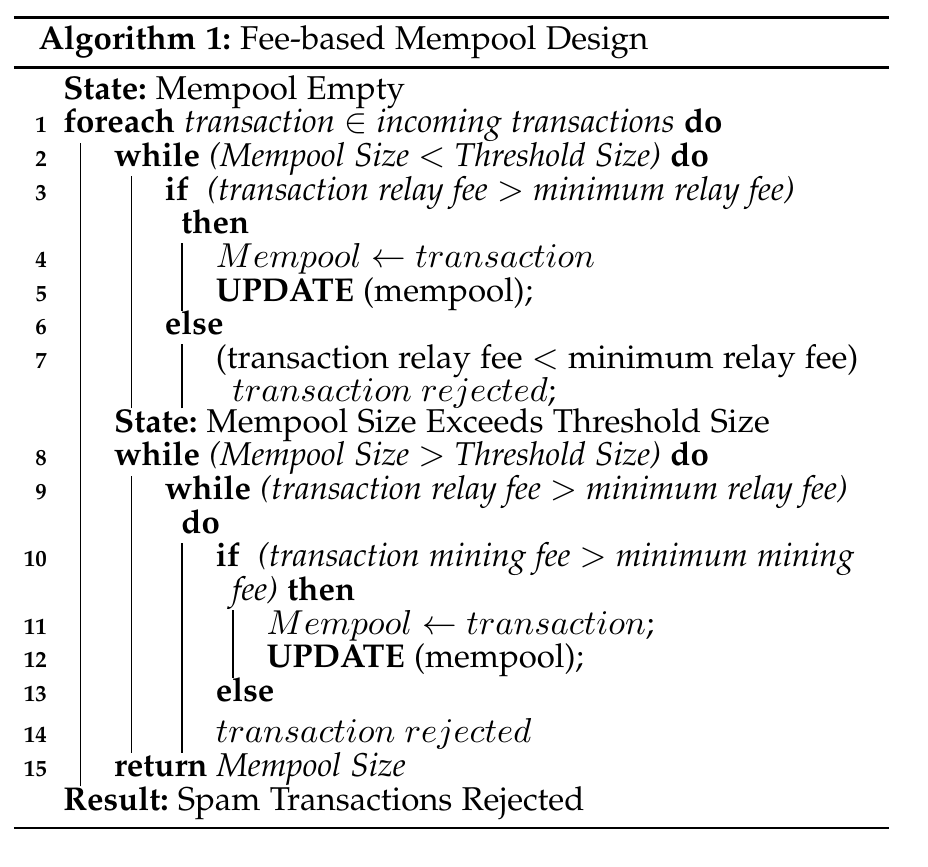}


\end{figure}

\begin{figure*}[!t]
\hfill
\begin{subfigure}[Confusion Matrix\label{fig:fee-tp}]{\includegraphics[width=5.5cm]{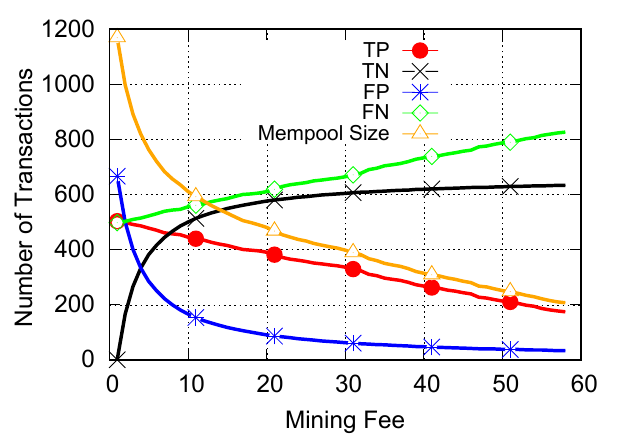}} 
\hfill
\end{subfigure}
\begin{subfigure}[Evaluation Parameters \label{fig:fee-precision}]{\includegraphics[width=5.5cm]{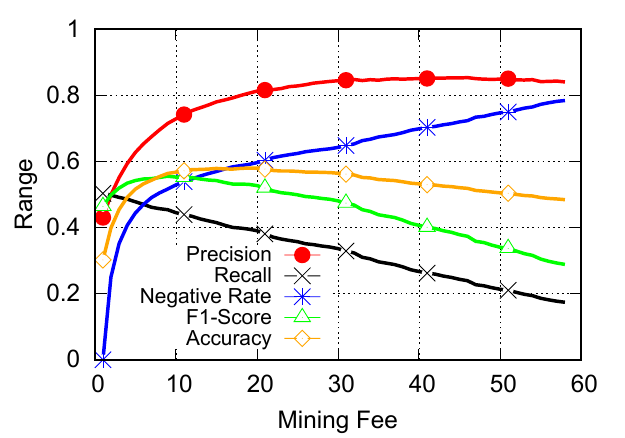}}
\hfill
\end{subfigure}
\begin{subfigure}[Accuracy, Size Ratio and Precision  \label{fig:feediff}]{\includegraphics[width=5.5cm]{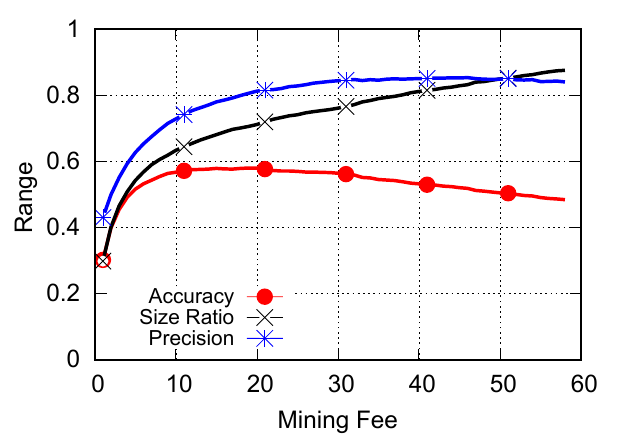}} 
\hfill
\end{subfigure}

\caption{Analysis of Fee-based Design. Notice that as the mining fee increases, the mempool size reduces. However increasing mining fee also affects benign transactions which is why the accuracy of detection decreases with increasing mining fee. An optimum cut-off fee can be selected from~\autoref{fig:feediff} based on the trade-off between accuracy and size ratio.  } 

\label{fig:fee-based}
\end{figure*}

\subsubsection{Analysis of \contraf} \label{sec:analysis_first}
In the following, we will analyze the workings of fee-based design and its utility in light of our threat model. We will limit the number of transactions an attacker can generate within his budget by increasing the mining fee threshold. We also observe how this design affects other benign users within the same network. 

In the current settings of Bitcoin, where an attacker only pays the relay fee to broadcast his transactions, if the mempools employ the fee-based design, all spam transactions will be rejected. As such, the mempool will only accept the transactions which pay both the relay fee and the mining fee. Legitimate users, on the other hand, will benefit by this design, since they will always pay the relay and the mining fee, so their transactions will be accepted. Once the attacker becomes aware of the fee-based design, the only way it can carry out the attack is by adapting to the new settings and masquerading as a benign user. The attacker can do that by adding mining fee to each transaction. Given a budget $B$, adding mining fee to each transaction will reduce the total number of transactions $T_a$ the attacker could generate in~\autoref{equation:total_transactions} will now become

\begin{equation} \label{equation:increasecost}
    T_a  = \frac{1024 \times B}{[(i \times k) + (o \times l) + i ] \times [R_f + M_f] \times T_\min }
\end{equation}

\begin{table}[t]
\centering
\caption{Confusion Matrix}
\scalebox{0.8}{
\begin{tabular}{l|l|c|c|c} 
\multicolumn{2}{c}{}&\multicolumn{2}{c}{Actual Transaction}&\\
\cline{3-4}
\multicolumn{2}{c|}{}&Legitimate&Malicious&\multicolumn{1}{c}{}\\
\cline{2-4}
 {Mempool }& Legitimate & TP & FP & \\
\cline{2-4}
{Transaction}& Malicious & FN & TN & \\
\cline{2-4}
\end{tabular}}
\label{table:confusion}
\end{table}

From~\autoref{equation:increasecost}, we can observe that the number of transactions the attacker can generate has an inverse relationship with the total fee paid per transactions. Using that relationship, we can adjust the fee parameter and investigate how it limits the attacker's capabilities. To do that, we simulate the affect of increasing the mining fee on the volume of transactions that the mempool accepts. We allocate a fixed budget to the attacker and select thresholds of minimum mining fee and maximum mining fee. Using~\autoref{equation:total_transactions}, we select a suitable budget for attacker that results into 1000 transactions with a minimum mining fee. Then, we generate 1000 benign transactions, each with a mining fee normally distributed over the range of the minimum and maximum mining fee. Using a discrete-event time simulation, we increase the mining fee and monitor its affect on transactions of the attacker and the benign users. 

We plot the results in~\autoref{fig:fee-based}, and use the confusion matrix in~\autoref{table:confusion} to evaluate the effect of the fee-based design on the mempool. We classify the true positives and the false positives as benign and malicious transactions accepted by the mempool respectively. We classify the false negatives and the true negatives as benign and malicious transactions rejected by the mempool. We plot the results of confusion matrix in~\autoref{fig:fee-tp}. The results show that with the increase in the mining fee threshold, the mempool size (TP+FP), malicious transactions (FP) and benign transactions (TP) decrease. The trend of (FP) is explained by~\autoref{equation:increasecost}. With a fixed budget, increasing the mining fee decreases the total transactions $T_a$. Accordingly, the size of the mempool also decreases due to fewer spam transactions (FP). However, increasing the mining fee also limits the fee-paying benign users. This, in turn, explains the trend of decreasing (TP).

\subsubsection{Evaluation Results} \label{sec:evalres}
From~\autoref{fig:fee-tp} and the evaluation criteria defined above, we measured the precision and the accuracy of our design. From ~\autoref{fig:fee-precision}, we observed that the accuracy increases with the mining fee to a maximum value and then decreases. From \autoref{fig:fee-precision}, we found the minimum fee cutoff corresponding to the maximum accuracy.

In~\autoref{fig:feediff}, we the plot accuracy and size ratio; the size ratio is the fraction of mempool transactions out of the total number of incoming transactions where a lower size ratio indicates higher size optimization. The results in~\autoref{fig:feediff} show that at a fee threshold of 13, we achieve $60\%$ accuracy, $70\%$ size optimization, and $78\%$ precision. Increasing the fee parameter further increases the size optimization but decreases the accuracy. Therefore, the fee-based design presents a trade-off between the size efficiency and the accurate detection of malicious transactions.

\subsubsection{Shortcomings of \contraf} \label{section:feelimitations}
To understand the limitations of ``Fee-based Mempool Design'' we highlight the nature of some transactions in Bitcoin. Suppose Alice sends 10 BTC to Bob in a transaction. That transaction is yet to be verified and mined, but Bob spends them by sending 5 BTC to Charlie. For Bob's transaction to be successfully mined, its parent transaction by Alice needs to be mined first. This sequence of transactions is known as parent-child transaction \cite{parent-child}. For a child transaction to become benign, its parent transaction needs to be mined first. Oftentimes when priority factor of a parent transaction is low, the child transaction increases the mining fee to increase the overall priority factor. This process is called ``Child Pays For Parent'' (CPFP). For benign users, this situation might be undesirable, since more child transactions can lead to transactions getting stuck. However, the same situation can be viewed as an opportunity by the attacker to circumvent the fee-based design and carry out the same attack at a lower cost.

For transactions made in the attack phase, their parent transactions in the distribution phase need to be verified and mined. The attacker can minimize the probability of transaction acceptance in the first phase by reducing their priority factor; e.g., by paying a minimum relay fee and no mining fee. Once the parent transactions have a lower probability of acceptance in the first phase, the child transactions can increase their priority factor by increasing the relay and mining fees. As a result, and when the mempools apply the fee-based countermeasures, spam transactions of the attack phase will get in mempool. After the mempool's size reaches the baseline threshold, the mempool will check for the incoming transactions with the minimum relay and mining fee. Since now the transactions of sybil accounts will pay both the relay and the mining fee, therefore, they will be accepted and the attacker will succeed.

\BfPara{Countermeasure} One way to address this problem is to prioritize the incoming transactions on the basis of mining fee. Mempool can sort the incoming transactions for the fee value and accept the ones which pay higher fee. As we increase the mining fee, the capability of attacker to produce transactions reduces~\autoref{equation:increasecost}. The attacker is constrained by the budget and increasing mining fee reduces the number of transactions he can produced. We can observe this trend in~\autoref{fig:fee-tp}. Although this reduces the number of spam transactions in the mempool and optimizes its size, it also reduces accuracy and the number of benign transactions that get accepted. As the fee parameter is increased, the capability of all the benign users to pay higher fee also decreases. To this end, the fee-based countermeasures do limit the attacker from flooding the mempool, but they also limit the number of benign transactions that successfully pass the fee threshold. To address these limitations we propose age-based countermeasures.

\subsection{\contraa: Age-based Countermeasures}

\subsubsection{\contraa{}'s Design} \label{sec:second_policy}
To limit attacker's chances of success, we propose the ``Age-based Mempool Design'' which addresses the limitations of our previous model. For this design, we leverage the confirmation factor or ``age'' of a transaction to distinguish between benign and malicious transactions. In Bitcoin, the age of a transaction determines how many block confirmations it has achieved over time (\textsection\ref{parg:confirmation}).

For this design in Algorithm 2, we assume that the baseline size threshold of the mempool has been reached, and the mempool is only accepting transactions which are paying the relay fee as well as the mining fee. Now, for each incoming transaction, we count the number of inputs or parent transactions. We initialize a variable ``average age'' and set its value to 0. Next, we calculate the average age of the transaction by adding the age of each parent transaction and dividing by the total number of parent transactions. This gives an estimate of confirmation score of the incoming transaction. Then, we apply a ``minimum age limit'' filter on the mempool. The ``minimum age limit'' can take any arbitrary value greater than 0. According to Bitcoin Developers Guide \cite{developer}, a confirmation score of 6 is considered good for any transaction. If the transaction's mean age value fulfills the age criteria, only then the mempool accepts the transaction. 

A transaction in Bitcoin has an input pointer to the spendable transaction  it previously received. For spam transactions, these inputs are not spendable and are less likely to be mined, serving the objectives of the attacker who intends to broadcast spam transactions which eventually get rejected. Although the age factor is taken into account for transactions, it is not considered while broadcasting those transactions. As such, attackers may exploit this feature of the system by broadcasting spam transactions and flooding the mempools without losing bitcoins. In this design, we apply the check on the age of the incoming transactions. In the attack phase (\textsection\ref{sec:secondphase}), the spam transactions will have input pointers of a parent transaction that will not be confirmed in any block. The age of parent transactions from the distribution phase will be 0.

\begin{figure*}[!t]
\hfill
\begin{subfigure}[Confusion Matrix\label{fig:age-tp}]{\includegraphics[width=5.5cm]{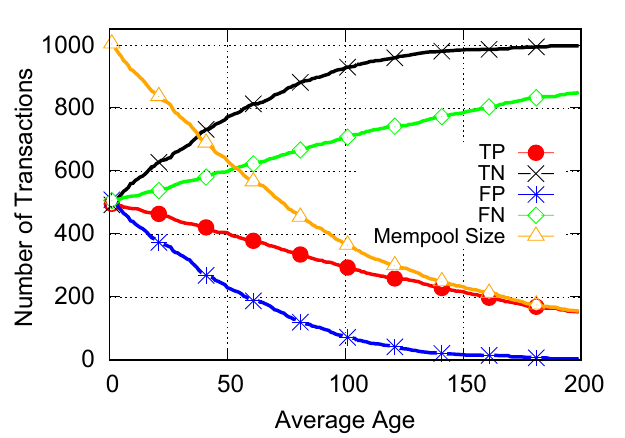}} 
\hfill
\end{subfigure}
\begin{subfigure}[Evaluation Parameters \label{fig:age-pr}]{\includegraphics[width=5.5cm]{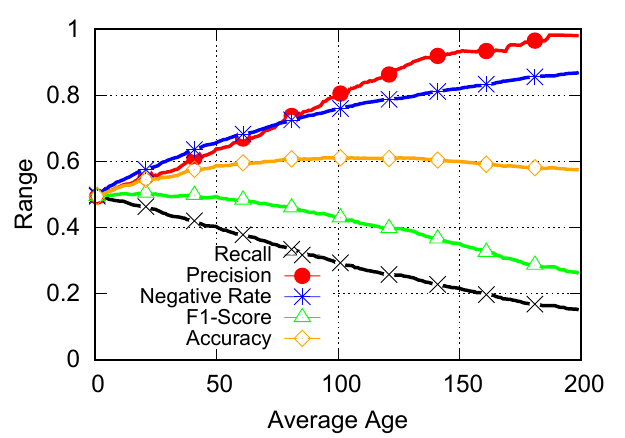}}
\hfill
\end{subfigure}
\begin{subfigure}[Accuracy, Size Ratio and Precision  \label{fig:agediff}]{\includegraphics[width=5.5cm]{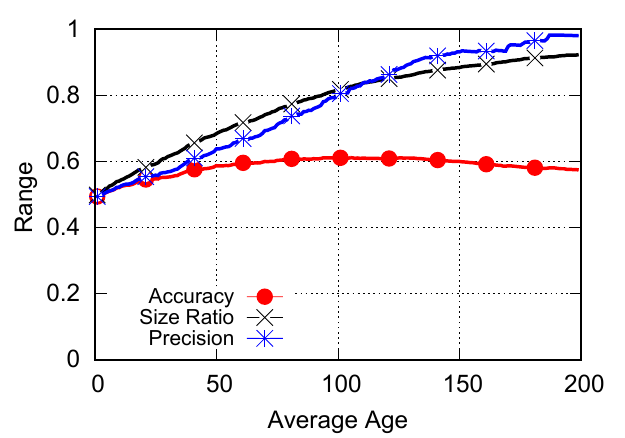}} 
\hfill
\end{subfigure}
\caption{Analysis of Age-based Design. Notice that with age-based design, the accuracy, precision, and size ratio are comparatively higher than the fee-based design. This policy is effective in rejecting the unconfirmed transactions.} 
\label{fig:age-based}
\end{figure*}

Using this knowledge about the nature of spam transactions, we compute the average age of all the input pointers (parent transactions); minimum age value of 1 means that all transactions coming into the pool are confirmed in at least the most recent block of the blockchain. In Bitcoin, once the transaction is mined into a blockchain, its parent transaction is removed from the ``UTXO'' set and cannot be spent again. The transaction itself becomes the new spendable ``UTXO.'' With these advantages, the age-based design prevents the system from spam transactions.

Once this design is implemented, if a user tries to spend his coins, he needs to have at least one valid confirmation for his transaction. This gives advantage to the benign user who can make a normal transaction with a confirmed parent transaction of significant age. On the other hand, the attacker's spam transactions will be rejected due to low confirmation factor despite paying high mining fee.

\begin{figure}
\centering
\includegraphics[width=0.48\textwidth]{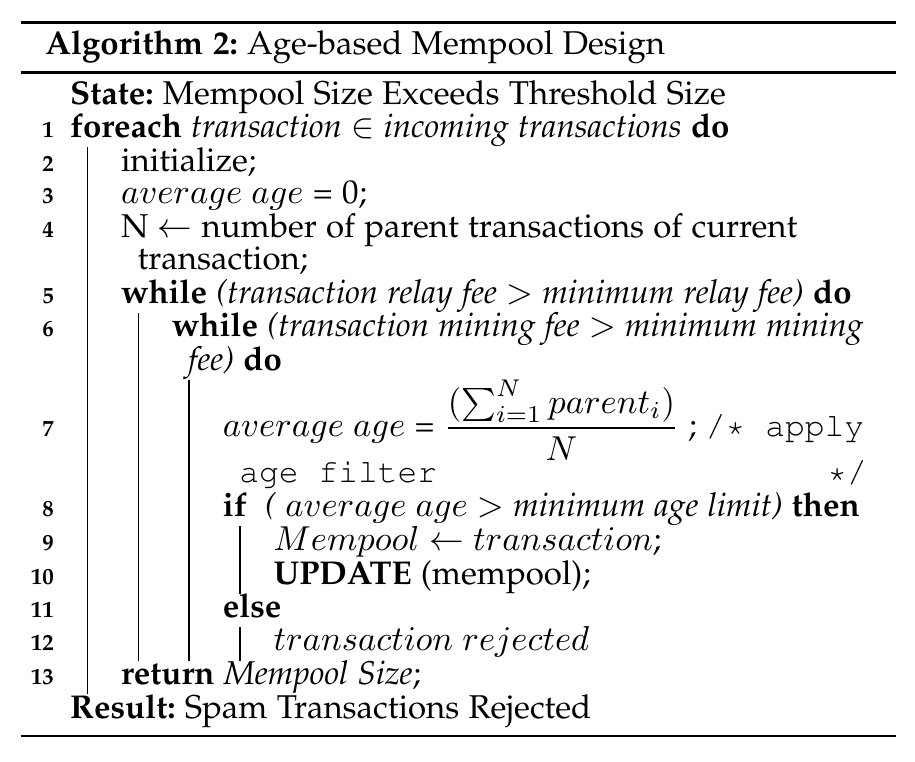}
\end{figure}

\subsubsection{Analysis of \contraa} \label{sec:analysis_second}
In this section, we analyze the working of ``Age-based Mempool Design'' and how it can help in countering DoS attack. For this design, we have established that the attacker has the capability of circumventing the ``Fee-based design'' and is willing to pay the relay fee and the mining fee in all of his transactions. Also, the attacker knows that his transactions will not be verified, so he pays higher relay and mining fee than the benign users.

We carried out our second experiment to analyze the working of the age-based countermeasures. We set a minimum age limit and a maximum age limit as thresholds for the incoming transactions. For the attacker, the only set of transactions with age value greater than 1 are generated in the distribution phase. Child transactions made in the attack phase were assigned 0 age value due to unconfirmed parent transactions. To capture that, we normally distribute the average age value of all malicious transactions from 0 to the minimum age limit. The average age value of all benign transactions was set from 0 to the maximum age limit. A total of 2000 transactions were generated with half of them being malicious and half being benign. Then, we applied the age-based design on all the incoming transactions at the mempool. We increased the age requirement for the transactions and evaluated the detection accuracy and the mempool state for each transaction.

\subsubsection{Evaluation Results}
Using the same confusion matrix in~\autoref{table:confusion} and evaluation parameters in (\textsection\ref{sec:evalpm}). The results in~\autoref{fig:age-based} show that upon increasing the average age the malicious transactions (FP) decrease sharply. The mempool size decreases to a point where there are only benign transactions left in the mempool. Due to low (FP) and higher (TP), the precision reaches a close to 1~\autoref{fig:age-pr}. In~\autoref{fig:agediff}, it can be observed that at an average age value of 100 we achieve $60\%$ accuracy, $80\%$ size optimization and $80\%$ precision. As we increase the age parameter to 200, the accuracy does not decrease like the fee-based design while the size ratio increases up to $90\%$ and the precision increases up to $98\%$. This shows that the age-based prevents a majority of malicious transactions from entering into the pool and helps the benign users in getting their transactions accepted.

In these settings, if the attacker intends to spam the network, he needs to have majority of his transactions confirmed in the blockchain. However, in our attack model, we have described that confirmation is undesirable for the attacker since it results in losing bitcoins in mining and relay fee. In Bitcoin, recall that the average block mining time is 10 minutes. For a single confirmation of all of the transactions, the attacker has to wait on average for 10 minutes. Using the results from~\autoref{fig:agediff}, the attacker will have to wait a minimum of 100 blocks to relaunch the attack. With average block computation time of 10 minutes, 100 blocks lead to 16 hours of delay. Even if the attacker still plans to carry out the attack after waiting and paying all the fee, he will not be able to flood the mempool. The best the attacker might achieve will be occasional network stressing with series of transactions; higher attack cost and low incentive will discourage the attacker. Therefore, the age-based design offers more security against the DoS attacks without affecting the benign users. 

\subsubsection{Shortcomings of \contraa}
Although the age-based countermeasures provide an effective defence in mitigating DoS attack on Bitcoin, there are some limitations in this design. Primarily, it requires all the incoming transactions to have a confirmed parent transaction.The average mining time for a block in Bitcoin is 10 minutes \cite{developer}. Depending on the bottleneck and size of the mempool, transaction verification can take even longer. In fast transactions where users cannot wait for verification, their transactions will be rejected by the mempool. An illustration of the fast transaction is bitcoin accepting vending machine. However, we do not see Bitcoin evolving into such applications any soon, so we do not consider it as a significant problem.

\subsection{\contras: Size-based Countermeasures}\label{sec:size_policy}
As discussed in \autoref{sec:introduction}, the block size is a key bottleneck dictating the transaction throughput. The size limit of blocks enables the attacker to create a queue of pending transactions in the mempool. Ideally, if the block size is set equal to the mempool size, the mempool queue can be virtually eliminated. However, as discussed in~\cite{ZamaniM018}, increasing the block size may have multiple drawbacks, including: \cib{1} increasing the blockchain size overhead, \cib{2} increase the transmission and propagation delays in the network, and \cib{3} increase the processing overhead at the receiving node. Therefore, the block size cannot be set equal to the mempool size. However, we postulate that a calculated increase in the block size can prevent the mempool flooding without sacrificing the performance. Conti \etal \cite{ContiELR18} show that Bitcoin can process a block size of up to 8 MB with tolerable size overhead and delay. In response to that, the Bitcoin community experimented with two soft works called {\em Segwit} and {\em Segwit2x}, which allow the block size to exceed the standard threshold of 1 MB. Motivated by these changes, in this section we will investigate the usefulness of block size in countering the mempool flooding attack.

\begin{figure}[t]
	\centering
	\includegraphics[width=1\linewidth]{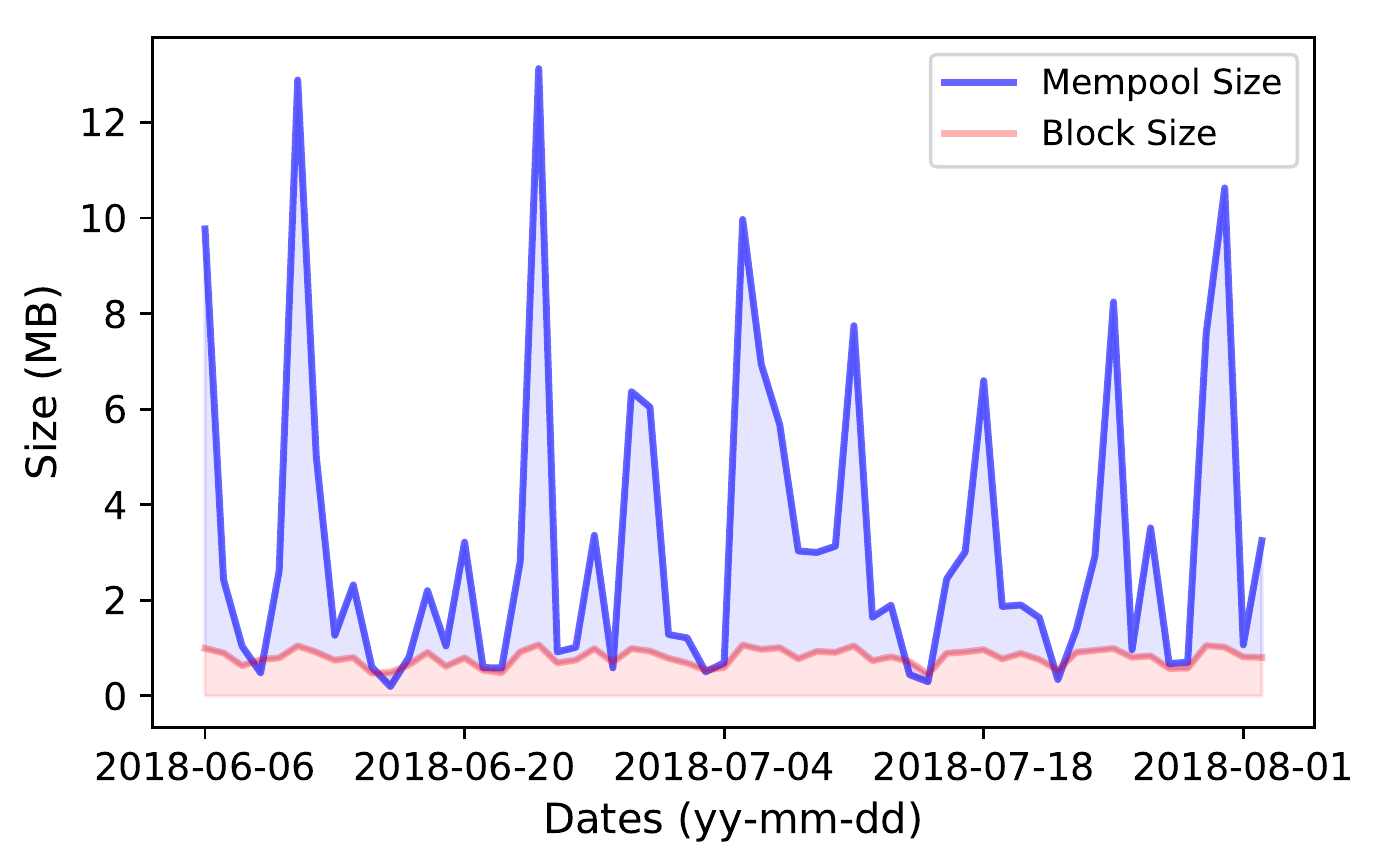}
	\caption{Difference between the Bitcoin block size and mempool size. The gap (shaded in blue color) provides advantage to the attacker to flood the mempool. Naturally, if the block size limit is increased, the gap can be reduced and the attack can be prevented.     }
	\label{fig:Size}
\end{figure}

\begin{figure}[t]
	\centering
	\includegraphics[width=1\linewidth]{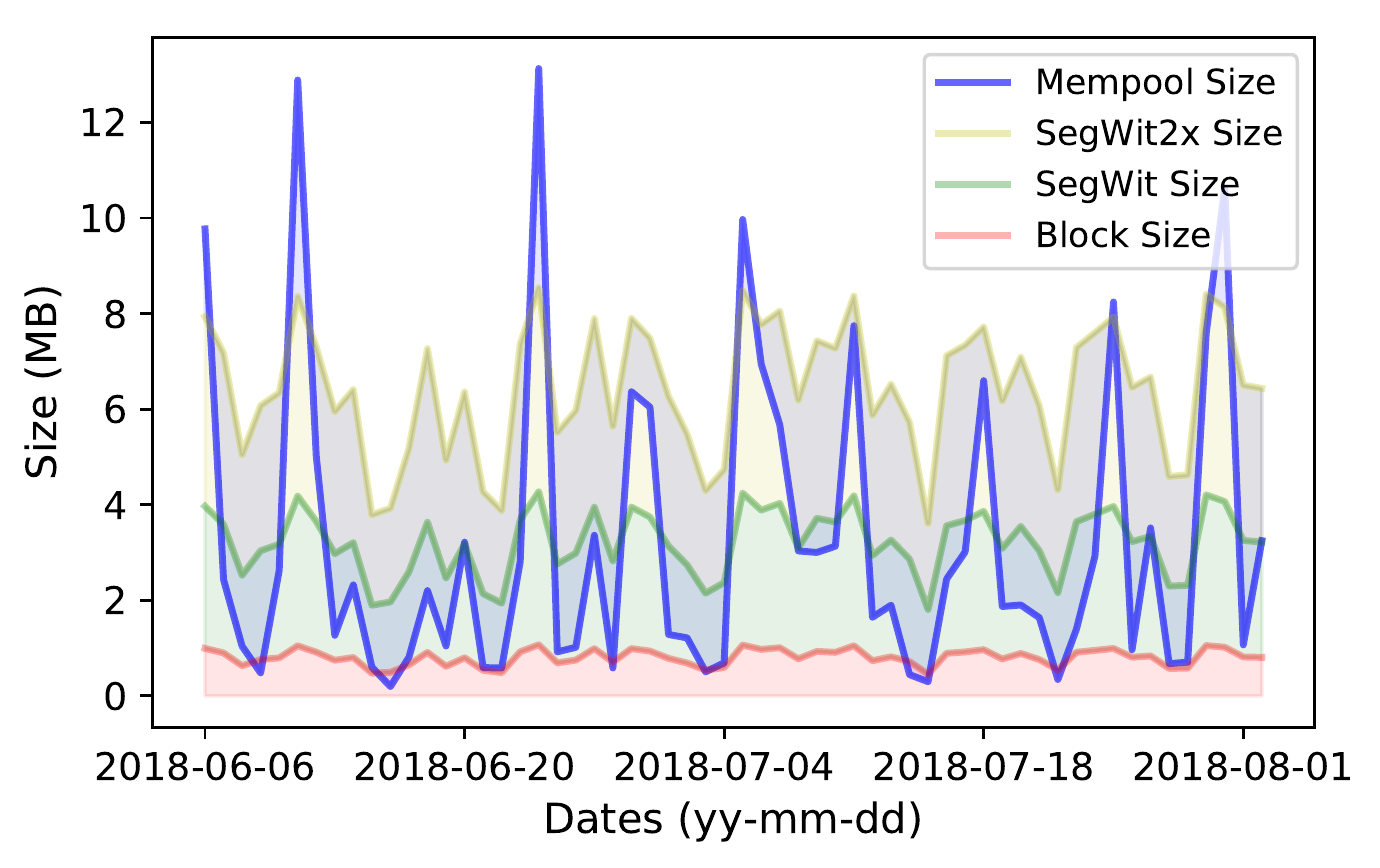}
	\caption{The effect of {\em Segwit} and {\em Segwit2x} on the mempool size. Notably, the {\em Segwit2x} minimizes the gap between the mempool size and the block size. As a result, during an attack, attacker's transactions will be mined in the blockchain.     }
	\label{fig:SizeTwo}
\end{figure}

Before analyzing the size-based design in detail, with the results shown in \autoref{fig:Size} we provide an intuitive overview of the gap between the block size and the mempool size. The blue region in~\autoref{fig:Size} shows the size of the pending transactions, and an attacker would ideally want to maintain the size in the blue zone to maintain the pressure on the mempool and avoid losing resources through transaction confirmation. In contrast, if the gap between block size and the mempool size is reduced, the transaction backlog will decrease, and the attacker will risk losing resources. Therefore, the block size variation can be modeled as a probability distribution that evaluates the chances of the attacker's transactions to be mined. The intuition behind this strategy is that if the mining probability is high, the attacker will be discouraged from launching the attack. Moreover, it will also reduce the gap between the block size and the mempool size, shown in \autoref{fig:Size}, thereby preventing the mempool flooding and the fee hike. Compared to the fee-based and age-based countermeasures, in this design we do not use transaction properties to isolate malicious transactions from benign transactions, and instead maximize the chances of transaction confirmation to raise the attack cost. In~\autoref{fig:SizeTwo}, we illustrate the effect of increasing the block size (as proposed by {\em SegWit} and {\em Segwit2x}) on the mempool size. Note that the blue region shown in \autoref{fig:Size} is significantly reduced in \autoref{fig:SizeTwo}. As a result, if the attacker has to flood the mempool he will have to generate transactions at a much higher rate with a high probability of losing balance.  

\subsubsection{Analysis of \contras} \label{sec:analysisSize}
For design analysis, we first define the system parameters. From \autoref{sec:mod}, we use the network throughput $b[t]$, the mempool size $Q$, and the incoming transactions rate $a[t]$. Moreover, we define the block size as $\mathcal{S}$, the set of transaction in the mempool as $\mathcal{T}$, the attacker's transactions in $\mathcal{T}$ as $\mathcal{I}$, the priority of a transaction $t_{i} \in \mathcal{T}$ as $\mathcal{P}(i)$, the mining probability of a transaction as $P(i)$, and the priority factor of a transaction as $\gamma_{i}$. The mempool state is defined as: 
\begin{align} \label{eq:decen}
    \text{Mempool State} &= \begin{cases} 
{a[t] < b[t]} & {Q = 0} \text{ Empty}\\
{a[t] > b[t]} & {Q \textgreater}{ 0} \text{ Attack}
\end{cases}
\end{align}

Next, we define the priority factor $\gamma_{i}$ for a transaction $t_{i} \in \mathcal{T}$. In cryptocurrencies, the priority is given to a transaction that pays a higher mining fee. Let $f_{1},\cdots,f_{v}$ be the fee paid by each transaction, where $v = |\mathcal{T}|$ is the total number of transactions in $\mathcal{T}$. We compute a normalized value for $\gamma_{i}$ and the priority factor $\mathcal{P}_{i}$  as follows: 
\begin{align} 
\label{eq:pri}
\gamma_{i} &= \frac{f_{i} - \min(f_{1}...f_{v}) }{\max(f_{1}...f_{v}) - \min(f_{1}...f_{v})}, 
\mathcal{P}(i) = \frac{\mathcal{S}}{Q} \times \gamma_{i}
\end{align} 

From the transaction priority and the size difference between the block and the mempool, we compute the transaction confirmation probability $P(i)$ as: 
\begin{align} 
\label{eq:pri2}
P(i) &= {\mathcal{P}(i)}\Biggm/{\sum\limits_{z=1}^{|\mathcal{T}|} \mathcal{P}(z)}.
\end{align} 

Following from \autoref{eq:pri2}, we define the aggregate probability $P(j)$ for all of the attacker's transaction in $\mathcal{I}$ as:
\begin{align} 
\label{eq:pri3}
P(j) &= {\sum\limits_{y=1}^{|\mathcal{I}|} \mathcal{P}(y)}\Biggm/{\sum\limits_{z=1}^{|\mathcal{T}|} \mathcal{P}(z)}
\end{align} 

From~\autoref{eq:pri}, we notice that if all transactions pay the same fee then the priority factor of any transaction will be 0. As a result, $P(i)$ will be undefined. However, the likelihood of such an event is small, since we mention in \autoref{sec:ddosmem} that the attacker will try to pay less fee than the mining fee of the benign users, facilitated by the lack of standard fees in Bitcoin. With $n$ users in Bitcoin, and assuming that each user generates one transaction, the probability that all users generate the same transaction fee is $\frac{1}{n}$. Since there are more than 1 million Bitcoin users \cite{statista19}, the probability for such an event is negligible.

When the size-based design is applied, a miner first checks the mempool state, shown in \autoref{eq:decen}. If the queue is empty, then the newly mined block size will be restricted to the standard size. However, when $a[t] > b[t]$, the miner would increase the block size. As a result, $Pr(i)$ for each transaction will also increase. Our goal is to evaluate the mining probability for an attacker when the block size is increased. Intuitively, if the block size is large, the attacker will be discouraged from launching the attack, fearing balance exhaustion. Additionally, we assume that attacker only pays the minimum relay fee to flood the mempool. 

\subsubsection{Evaluation Results} \label{sec:evalsize}
\begin{figure}[t]
	\centering
	\includegraphics[width=1\linewidth]{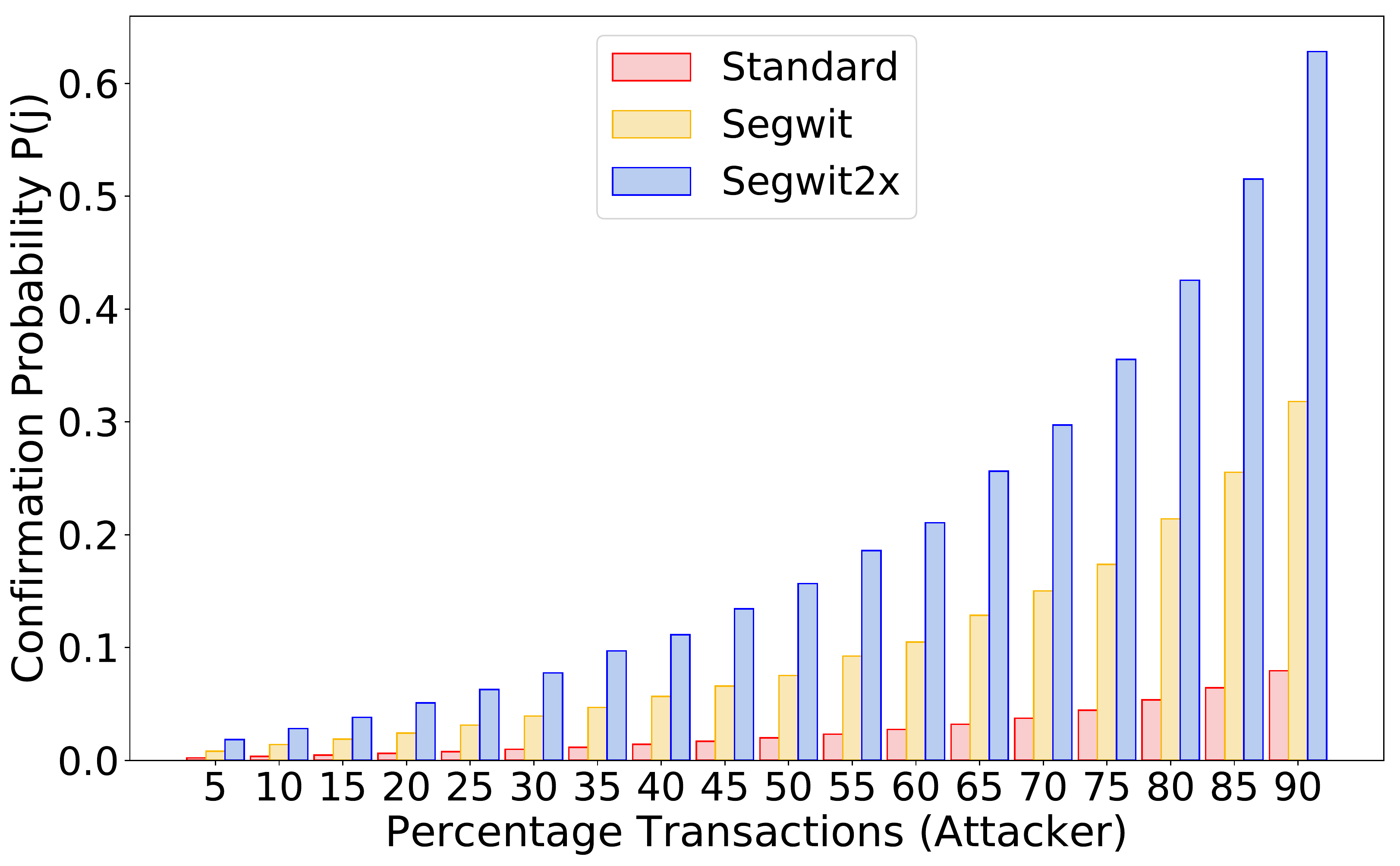}
	\caption{Results obtained from the simulation of the size-based policy. The x-axis show the percentage of the attacker's transaction among the total transactions. The y-axis is the probability of acceptance of all of the attacker's transactions $P(j)$. The results show that the increase in block size will increase the probability of transactions. Therefore, the attacker will risk losing his balance.    }
	\label{fig:SizeThree}
\end{figure}

To evaluate the effect of \contras, we generate a series of transactions, each of which is with a size of 2 KB. Each transaction was assigned a fee, distributed normally between minimum and maximum fee threshold (\autoref{sec:first_policy}). The mempool size is then the sum of the size of all transactions. Next, we start with fixed transaction size for the attacker (10\%), and sequentially increase the size to 90\% of the total transactions. For each of the attacker's transaction, we randomly select the transaction fee between 0 and the minimum relay fee. We apply the size-based policy by setting the block size to 1 MB (current block size in Bitcoin), 4 MB (the {\em segwit size}), and 8 MB ({\em segwit2x size}). We use the same discrete-event simulation settings as used in the earlier experiments. In our simulations, we evaluate the effect of the block size on the probability of transaction confirmation and report the results in \autoref{fig:SizeThree}. Our results show the effect of increasing the block size on the confirmation probability $P(j)$ of the attacker's transactions in $\mathcal{I}$ $\in$ $\mathcal{T}$. The general trend indicates that when the block size  increased, $P(j)$  also increased, which would naturally affect the attacker's balance. Moreover, it is worth noting that when the attacker increases the rate of the incoming transactions, shown on the x-axis, $P(j)$ increases exponentially. Therefore, the risk of balance exhaustion becomes more significant. 

Apart from the promising results achieved by \contras, we believe that  applying \contras in current Bitcoin protocol is relatively easier than \contraf{}'s and \contraa{}'s. The \contras will be applied on the block rather than the mempool. This would require minimal changes in the client software. Moreover, the size-based design specifies concrete parameters (4 MB or 8 MB), which prevent conflicting views. In the fee-based and age-based countermeasures, nodes can have different size of the mempool due to variance in the incoming transaction rate. As a result, they will apply different threshold parameters to control the size of the mempool, which can lead to conflicting states at each node. On the other hand, with concrete parameters for the block size, such a requirement to alter policies at the mempool will be alleviated. As part of our future work, we will analyze the asynchronous settings of Bitcoin network to understand the  disagreements in the mempool view. Using that, we will derive bounds on the fee-based and age-based design to prevent the mempool attack while preserving the agreement on the mempool state.

\section{Conclusion} \label{sec:conclusion}
In this paper, we identify a DoS attack on Bitcoin mempools that pushes the users into paying higher mining fees. Attacks on Bitcoin mempools have not been addressed previously, and we propose three countermeasures to the problem: fee-based, age-based, and size-based designs. Using  simulations and various analyses, we conclude that when the attack is not severe, the fee-based design is more effective in mempool size optimization. However, the fee-based design does so by affecting both the attacker and the benign users. When the attack is severe, the age-based design is more useful in helping benign users while discarding maximum spam transactions. However, both designs use features from incoming transactions to distinguish between malicious and benign transaction. Our third design overcomes this shortcoming by simply increasing the block size and limiting the attacker's capabilities. The strength of the size-based design lies in concrete policy parameters that can be easily applied in the blockchain system, without any major protocol modifications.

\bibliography{ref.bib}

\end{document}